\documentstyle[aps,pra,psfig]{revtex}  

\def\text#1{\mbox{#1}}

\def\psfig#1{}
\def\newpageAA{}

\begin{document}

\author{Vladimir N. Serkin$^{a}$ and Akira\ \ Hasegawa$^{b}$ \\
$^{a}$Benemerita Universidad Autonoma de Puebla, Instituto de Ciencias\\
Apdo Postal 502, 72001 Puebla, Pue., Mexico\\
$^{a}$General Physics Institute, Russian Academy of Science,\\
Vavilova 38, 117942 Moscow\\
$^{a}$email address: vserkin@hotmail.com\\
$^{b}$Kochi University of Technology and NTT Science and Core Technology\\
ATR BLDQ., 2-2 Hikaridai Seikacho Sorakugun\\
Kyoto, Japan 619-0288}
\date{15 February 2000}
\title{{\bf Femtosecond soliton amplification in nonlinear dispersive traps and
soliton dispersion management}}
\maketitle

\begin{abstract}
The nonlinear pulse propagation in an optical fibers with varying parameters
is investigated. The capture of moving in the frequency domain femtosecond
colored soliton by a dispersive trap formed in an amplifying fiber makes it
possible to accumulate an additional energy and to reduce significantly the
soliton pulse duration. Nonlinear dynamics of the chirped soliton pulses in
the dispersion managed systems is also investigated. The methodology
developed does provide a systematic way to generate
infinite ``ocean'' of
the chirped soliton solutions of
the nonlinear Schr\"odinger equation (NSE)
with varying coefficients.

{\bf Keywords and PACS numbers: }Femtosecond solitons amplification,
dispersion management

42.65 Tg, 42.81 Dp
\end{abstract}

\section{INTRODUCTION}

In 1973 Hasegawa and Tappert
\cite{HasegawaA73}
showed theoretically that an optical pulse
in a dielectric fibers forms an envelope solitons, and in 1980 Mollenauer,
Stolen and Gordon
\cite{MollenauerLF80}
demonstrated the effect experimentally. This discovery
is significant in its application to optical communications. Today the
optical soliton is regarded as an important alternative for the next
generation of high speed telecommunication systems.

The theory of NSE solitons was developed for the first time in 1971 by
Zakharov and Shabad \cite{ZakharovVE71}.
The concept of the soliton involves a large number
of interesting problems in applied mathematics since it is an exact
analytical solution of a nonlinear partial differential equations. The
theory of optical solitons described by the nonlinear Schr\"odinger equation
has produced perfect agreement between theory and experiment
\cite{TaylorJR92}.

In this paper we present mathematical description of solitary waves
propagation in a nonlinear dispersive medium with varying parameters.

The soliton spectral tunneling effect was theoretically predicted in
\cite{SerkinVN93}.
This is characterized in the spectral domain by the passage of a femtosecond
soliton through a potential barrier-like spectral inhomogeneity of the group
velocity dispersion (GVD), including the forbidden band of a positive GVD.
It is interesting to draw an analogy with quantum mechanics where the
solitons are considered to exhibit particle-like behavior. The soliton
spectral tunneling effect also can be considered as an example of the
dynamic dispersion soliton management technique. In the first part of the
paper we will concentrate on the problem of femtosecond solitons
amplification. We will show that spectral inhomogeneity of GVD allows one to
capture a soliton in a sort of spectral trap and to accumulate an additional
energy during the process of the soliton amplification. In the second part
we will consider the problem of the short soliton pulse propagation in the
nonlinear fiber with static non-uniform inhomogeneity of GVD. The
methodology developed does provide a systematic way to generate
infinite
``ocean'' of the chirped soliton solutions of NSE model with varying
coefficients.

\section{FEMTOSECOND\ SOLITON\ AMPLIFICATION}

It is well known that due to the Raman self-scattering effect
\cite{DianovEM85} (called
soliton self-frequency shift \cite{MitschkeFM86})
the central femtosecond soliton frequency
shifts to the red spectral region and so-called colored solitons are
generated. This effect decreases significantly the efficiency of resonant
amplification of femtosecond solitons. The mathematical model we consider
based on the modified NSE including the effects of molecular vibrations and
soliton amplification processes (see details in \cite{AfanasjevVV92}):
\begin{equation}
i{\frac{\partial \psi }{\partial z}}={\frac 12}{\frac{\partial ^2\psi }{%
\partial \tau ^2}+i\sigma }\frac{\partial ^3\Psi }{\partial \tau ^3}%
+(1-\beta )|\psi |^2\psi +\beta Q\psi +{\frac G2}P
\end{equation}

\begin{equation}
\mu ^2{\frac{\partial ^2Q}{\partial t^2}}+2\mu \delta {\frac{\partial Q}{%
\partial t}}+Q=|\psi |^2,
\quad\text{and},\quad
\gamma _a{\frac{\partial P}{%
\partial \tau }}+P(1+i\gamma _a\Delta \Omega )=i\psi ,
\end{equation}
As numerical experiments showed the GVD\ inhomogeneity as a potential well
allows one to capture a soliton in a sort of spectral trap. Figure 
\ref{fig01} shows
the nonlinear dynamics of the soliton spectral trapping effect in the
spectral domain. As soliton approaches the well, it does not slow down but
speeds up, and then, after it has got into the well, the soliton is trapped.
There exists a long time of soliton trapping in internal region of the well
.This effect opens a controlled possibility to increase the energy of a
soliton. As follows from our computer simulations the capture of moving in
the frequency space femtosecond colored soliton by a dispersive trap formed
in an amplifying optical fiber makes it possible to accumulate an additional
energy in the soliton dispersive trap and to reduce significantly the
soliton pulse duration.

\section{DISPERSION\ MANAGEMENT:\ CHIRPED\ SOLITONS}

Let us consider the propagation of a nonlinear pulse in the anomalous (or
normal) group velocity dispersion fiber of length $Z_1.$ The complex
amplitude $q$ of the light wave in a fiber with variable parameters $D_2(Z)$
, $N_2(Z)$ and $\Gamma (Z)$ is described by the nonlinear Schrodinger
equation
\begin{equation}
i\frac{\partial q}{\partial Z}+\frac 12D_2(Z)\frac{\partial ^2q}{\partial T^2%
}+N_2(Z)\mid q\mid ^2q=i\Gamma (Z)q
\end{equation}

{\bf Theorem 1}. Consider the NSE (3) with varying dispersion, nonlinearity
and gain. Suppose that Wronskian W[N$_2$,D$_2$] of the functions N$_2$(Z)
and D$_2$(Z) is nonvanishing, thus two functions N$_2$(Z) and D$_2$(Z) are
linearly independent. There are then infinite number of solutions of Eq. (3)
in the form of Eq.4
\begin{equation}
q(Z,T)=\sqrt{\frac{D_2(Z)}{N_2(Z)}}\text{ }P(Z)\text{ }Q\left[ P(Z)\cdot
T\right] \text{ }\exp \left[ i\frac{P(Z)}2\text{ }T^2+i\int%
\limits_0^ZK(Z^{^{\prime }})dZ^{^{\prime }}\right]
\end{equation}
where function $Q$ describes fundamental functional form of bright
$Q=\mbox{sech}(P(Z)T)$
or dark $Q=\mbox{th}(P(Z)T)$ NSE\ solitons and the real functions P(Z),
D$_2(Z)$, N$_2(Z)$
and $\Gamma (Z)$ are determined by the following nonlinear
system of equations :
\begin{equation}
\frac 1{P^2(Z)}\frac{\partial P(Z)}{\partial Z}+D_2(Z)=0\text{ ; }-\frac
12D_2(Z)P(Z)+\frac{W\left[ N_2(Z),D_2(Z)\right] }{2D_2(Z)N_2(Z)}=\Gamma (Z)
\end{equation}

{\bf Theorem 2}. Consider the NSE (3) with varying dispersion, nonlinearity
and gain. Suppose that Wronskian W[N$_2$,D$_2$] of the functions N$_2$(Z)
and D$_2$(Z) is vanishing, thus two functions N$_2$(Z) and D$_2$(Z) are
linearly dependent. There are then infinite number of solutions of Eq. (3)
of the following form Eq. 6
\begin{equation}
q(Z,T)=C\text{ }P(Z)\text{ }Q\left[ P(Z)\cdot T\right] \text{ }\exp \left[ i%
\frac{P(Z)}2\text{ }T^2+i\int\limits_0^ZK(Z^{^{\prime }})dZ^{^{\prime
}}\right]
\end{equation}
where function $Q$ describes the fundamental form of bright (or dark) NSE\
soliton and the real functions P(Z), D$_2(Z)$, N$_2(Z)$ and $\Gamma (Z)$ are
determined by the following nonlinear system of equations :
\begin{equation}
D_2(Z)=-\frac 1{P^2(Z)}\frac{\partial P(Z)}{\partial Z}\text{ ; }\Gamma
(Z)=\frac 12\frac 1P\frac{\partial P(Z)}{\partial Z}\text{ ; }%
N_2(Z)=D_2(Z)/C^2
\end{equation}
The function P(Z)\ is required only to be once-differentiable, but otherwise
arbitrary function, there is no restrictions.

To prove Theorems 1 and 2 we first construct a stationary localized solution
of Eq. (3) by introducing Kumar- Hasegawa's quasi-soliton concept
\cite{KumarS97,HasegawaA98,KodamaY98} 
through

\begin{equation}
q(Z,T)=\sqrt{\frac{D_2(Z)}{N_2(Z)}}\text{ }P(Z)\text{ }Q\left[ P(Z)\cdot
T\right] \text{ }\exp \left[ i\frac{P(Z)}2\text{ }T^2+i\int%
\limits_0^ZK(Z^{^{\prime }})dZ^{^{\prime }}\right]
\end{equation}
where $D_2$(Z), $N_2$(Z), $P(Z)$ and $K(Z)$ are the real functions of $Z.$
Substituting expression (8) into Eq. (3) and separating real and imaginary
parts we obtain the system of two equations
\begin{equation}
\frac 12\text{sign}(D_2)\frac{\partial ^2Q}{\partial S^2}
 +Q^3+\left( E-\frac{S^2}%
2\cdot \Omega ^2(Z)\right) \text{ }Q=0
\end{equation}

\begin{equation}
\frac{\partial P}{\partial Z}Q+P\frac{\partial Q}{\partial S}\frac{\partial S%
}{\partial Z}+\frac 12\frac 1{D_2(Z)}\frac{\partial D_2}{\partial Z}PQ-\frac
12\frac 1{N_2(Z)}\frac{\partial N_2}{\partial Z}PQ+\frac 12D_2P^2Q+D_2P^2T%
\frac{\partial Q}{\partial S}\frac{\partial S}{\partial T}=\Gamma PQ
\end{equation}
Where

\begin{equation}
S(Z,T)=P(Z)T\text{ ; }\frac{\partial S}{\partial Z}=T\frac{\partial P}{%
\partial Z}\text{ ; }\frac{\partial S}{\partial T}=P(Z)
\end{equation}
In Eq. (9) the parameters $E$ and $\Omega $ are 'the energy' and 'frequency'
of ordinary quantum mechanical harmonic ocsillator
\begin{equation}
\Omega ^2(Z)=\frac{D_2^{-1}(Z)}{P^2(Z)}\left( \frac 1{P^2(Z)}\frac{\partial P%
}{\partial Z}+D_2(Z)\right) \text{ ; }E(Z)=-K(Z)/P^2(Z)/D_2(Z)
\end{equation}
Eq. (9) represents the nonlinear Schrodinger equation for the harmonic
ocsillator. As must be in the case of Hamiltonian system Eq. (9) may be
written in the form
\begin{equation}
\frac{\delta H}{\delta Q^{*}}=0
\end{equation}

\begin{equation}
H=\int \text{ }\left[ \frac 12\text{sign}(D_2)
\left| \frac{\partial Q}{\partial X}%
\right| ^2+\frac 12\alpha \left| Q\right| ^4+\left( E-\frac{X^2}2\cdot
\Omega ^2(Z)\right) \text{ }\left| Q\right| ^2\right] dX
\end{equation}
The derivative in (13) is functional derivative. For the first time this
equation was solved numerically by Kumar and Hasegawa in \cite{KumarS97}
and gave rise a
new concept of quasi-solitons \cite{HasegawaA98,KodamaY98}.
Now we make the important assumption
about the solution of Eq. (9).

Let us consider the complete nonlinear regime when Eq. (9) represents the
ideal NLS\ equation, i.e. we will allow $\Omega (Z)\equiv 0$ , then from
(12) follows that
\begin{equation}
\frac 1{P^2(Z)}\frac{\partial P(Z)}{\partial Z}+D_2(Z)=0
\end{equation}
We now look for a solution of Eq. (10) which satisfies the condition (15).
Substituting the expression (15) and relations (11) into Eq. (10) we obtain
\begin{equation}
-\frac 12D_2(Z)P(Z)+\frac 12\frac 1{D_2(Z)}\frac{\partial D_2(Z)}{\partial Z}%
-\frac 12\frac 1{N_2(Z)}\frac{\partial N_2(Z)}{\partial Z}=\Gamma (Z)
\end{equation}
Using notation

\[
W\left\{ N_2,D_2\right\} =N_2\frac{\partial D_2(Z)}{\partial Z}-D_2\frac{%
\partial N_2(Z)}{\partial Z}
\]
one can obtain the soliton solution of Eq. 3 in the form of the chirped
solitons Eqs. 4-5 and Eqs. 6-7.. Consequently, we have found the infinite
''ocean'' of solutions. The methodology developed does provide a systematic
way of new and new chirped soliton solutions generation.

\section{DIFFERENT\ REGIMES OF\ SOLITON\ MANAGEMENT}

{\bf Lemma 1}: Soliton GVD\ management. Consider the NSE (3) with constant
nonlinear coefficient N$_2=const$ and with varying along Z-coordinate GVD\
parameter. Suppose that dispersion management function is known arbitrary
analytical function :D$_2$(Z)=$\Phi (Z)$ . The function $\Phi (Z)$\ is
required only to be once-differentiable and once integrable, but otherwise
arbitrary function, there is no restrictions. There are then infinite number
of solutions of Eq. (3) of the form of the chirped dispersion managed dark
and bright solitons Eq. 4, where the main functions P and $\Gamma $ are
given by
\begin{equation}
D_2(Z)=\Phi (Z)\text{ };\text{ }P(Z)=-\frac 1{\left[ C-\int \Phi
(Z)dZ\right] }
\end{equation}

\begin{equation}
\Gamma (Z)=\frac 12\frac{\Phi (Z)}{\left[ C-\int \Phi (Z)dZ\right] }+\frac
12\frac 1{\Phi (Z)}\frac{\partial \Phi (Z)}{\partial Z}
\end{equation}

{\bf Lemma 2:} Soliton intensity management. Consider the NSE (3) with
constant nonlinear coefficient N$_2=const$ and with varying along
Z-coordinate the dispersion and gain. Suppose that intensity of the soliton
pulse is determined by the known management function: D$_2$(Z)P$^2$(Z)=$%
\Theta (Z),$where the function $\Theta (Z)$\ is required only to be
once-differentiable and once integrable, but otherwise arbitrary function,
there is no restrictions. There are then infinite number of solutions of Eq.
(3) of the form of the chirped dispersion managed dark and bright solitons
Eq. 4 with parameters given by
\[
D_2(Z)P^2(Z)=\Theta (Z)\text{ ; }P(Z)=-\int \Theta (Z)dZ+C\text{ ; }D_2(Z)=%
\frac{\Theta (Z)}{\left[ C-\int \Theta (Z)dZ\right] ^2}
\]

\begin{equation}
\Gamma (Z)=\frac 12\frac{\Theta (Z)}{\left[ C-\int \Theta (Z)dZ\right] }%
+\frac 12\frac 1{\Theta (Z)}\frac{\partial \Theta (Z)}{\partial Z}
\end{equation}

{\bf Lemma 3:} Soliton pulse duration management: optimal soliton
compression. Consider the NSE (3) with constant nonlinear coefficient N$%
_2=const$ and with varying along Z-coordinate the dispersion and gain
coefficients. Suppose that pulse duration of a soliton is determined by the
known analytical function: P(Z)=$\Upsilon (Z)$, where the function $\Upsilon
(Z)$\ is required only to be two-differentiable , but otherwise arbitrary
function, there is no restrictions. There are then infinite number of
solutions of Eq. (3) of the form of the chirped dispersion managed dark and
bright solitons Eq. 4 with the main parameters given by
\begin{equation}
D_2(Z)=-\frac 1{\Upsilon ^2(Z)}\frac{\partial \Upsilon (Z)}{\partial Z}\text{
; }\Gamma (Z)=\frac 12\left( \frac{\partial \Upsilon (Z)}{\partial Z}\right)
^{-1}\frac \partial {\partial Z}\left( \frac 1{\Upsilon (Z)}\frac{\partial
\Upsilon (Z)}{\partial Z}\right)
\end{equation}

{\bf Lemma 4}: Soliton amplification management: optimal soliton
compression. Consider the NSE (3) with constant nonlinear coefficient N$%
_2=const$ and with varying along Z-coordinate the dispersion and gain
coefficients. Suppose that the gain coefficient is determined by the known
control function: $\Gamma $(Z)=$\Lambda (Z),$ where the function $\Lambda (Z)
$\ is required only to be once integrable , but otherwise arbitrary
function, there is no restrictions. There are then infinite number of
solutions of Eq. (3) of the form of the chirped dispersion managed dark and
bright solitons of the Eq. 4 where
\begin{equation}
\left| P(Z)\right| =\exp \left[ \int \exp \left( \int 2\Lambda (Z^{^{\prime
\prime }})dZ^{^{\prime \prime }}\right) dZ^{^{\prime }}\right]
\end{equation}

\begin{equation}
\left| D_2(Z)\right| =\frac{\exp \left( \int 2\Lambda (Z)dZ\right) }{\exp
\left[ \int \exp \left( \int 2\Lambda (Z^{^{\prime \prime }})dZ^{^{\prime
\prime }}\right) dZ^{^{\prime }}\right] }
\end{equation}

{\bf Lemma 5:} Combined dispersion and nonlinear soliton management.
Consider the NSE (3) with varying nonlinear coefficient N$_2(Z)$ and with
varying along Z-coordinate the dispersion and gain coefficients too. Suppose
that Wronskian W[N$_2$,D$_2$] is vanishing, or that the functions N$_2$(Z)\
and D$_2$(Z)\ are linearly dependent. Suppose also that the function D$_2$%
(Z) is determined by the initial control function D$_2$(Z)=$\Xi (Z),$where
the function $\Xi (Z)$\ is required only to be once integrable, but
otherwise arbitrary function, there is no restrictions. There are then
infinite number of solutions of Eq. (3) of the form of the chirped
dispersion managed dark and bright solitons of the Eq. 6 where
\begin{equation}
P(Z)=-\frac 1{\left[ C-\int \Xi (Z)dZ\right] }\text{ ; }N_2(Z)=D_2(Z)/C^2
\end{equation}

\begin{equation}
\Gamma (Z)=\frac 12\frac{\Xi (Z)}{\left[ C-\int \Xi (Z)dZ\right] }
\end{equation}
The analytical solutions for the different regimes of the main soliton
parameters management (intensity, pulse duration, amplification or
absorption ) in the case of W[N$_2$,D$_2$]=0 can be obtained by using
theorem 2.

Let us consider some examples. The case of $\Gamma (Z)\equiv 0$ and N$_2$%
(Z)=N$_2(0)$ corresponds to the problem of ideal GVD\ soliton management.
The soliton solution in this case is:
\begin{eqnarray}
q(Z,T) &=&-\eta N_2^{-1/2}(0)\exp (\text{ }\frac C2Z\text{ })\text{ sech }%
\left[ \eta T\exp (CZ)\right] \ \  \\
&&\exp \left[ -iT^2\frac C2\exp (CZ)-i\frac 12\eta ^2Z\exp (CZ)\right]
\end{eqnarray}

\begin{eqnarray}
q(Z,T) &=&\eta N_2^{-1/2}(0)\exp (\text{ }\frac C2Z\text{ })\text{ th }%
\left[ \eta T\exp (CZ\right]  \\
&&\ \ \ \exp \left[ iT^2\frac C2\exp (CZ)-i\eta ^2Z\exp (CZ)\right]
\end{eqnarray}
Here $T$ and $Z$ are ordinary variables and $C$ is arbitrary constant. If we
use the expressions D$_2$(Z)=constant and N$_2$=N$_2(0)$ then we obtain the
following solutions of Eq. (3) in the form of hyperbolically growing ideal
bright and dark solitons (for the first time reported in
\cite{MooresJD96,KhasilevVY96} 
\begin{equation}
q(Z,T)=-\frac{\chi \text{ }N_2^{-1/2}(0)}{(1-2\Gamma (0)Z)}\text{ sech}%
\left[ \frac{\chi T}{(1-2\Gamma (0)Z)}\right] \ \ \exp \left[ -i\frac{%
T^2\Gamma (0)}{(1-2\Gamma (0)Z)}-i\frac{\chi ^2Z}{2(1-2\Gamma (0)Z)}\right]
\end{equation}

\begin{equation}
q(Z,T)=\frac{\chi \text{ }N_2^{-1/2}(0)}{(1-2\Gamma (0)Z)}\text{ th}\left[
\frac{\chi T}{(1-2\Gamma (0)Z)}\right] \exp \left[ i\frac{T^2\Gamma (0)}{%
(1-2\Gamma (0)Z)}-i\frac{\chi ^2Z}{(1-2\Gamma (0)Z)}\right]
\end{equation}
In the case of $\Gamma (Z)\equiv G_0$ and N$_2$=$N_2(0)$ the solution of Eq.
3 is given by:
\begin{equation}
\text{ }Q(P(Z)T)=\eta N_2^{-1/2}(0)\text{ sech }\left[ \eta P(Z)T\right]
\end{equation}

\begin{equation}
\text{ }Q(P(Z)T)=\eta N_2^{-1/2}(0)\text{ th }\left[ \eta P(Z)T\right]
\end{equation}

\begin{equation}
P(Z)=-P(0)\exp (\frac 1{2G_0}(\exp (2G_0Z)-1))
\end{equation}

\begin{equation}
D_2(Z)=D_2(0)\exp (2G_0Z-\frac 1{2G_0}(\exp (2G_0Z)-1))
\end{equation}

When GVD is a hyperbolically decreasing function of Z
\begin{equation}
D_2(Z)=\frac 1{1+\beta Z}
\end{equation}
then from Lemma 1 follows the explicit soliton solution in the form of Eq. 4
\begin{equation}
P(Z)=-\frac 1{1-\frac 1\beta \ln (1+\beta Z)}
\end{equation}

\begin{equation}
\Gamma (Z)=\frac 1{2(1+\beta Z)}\left[ \frac{1-\ln (1+\beta Z)}{1-\frac
1\beta \ln (1+\beta Z)}\right]
\end{equation}

Let us consider the soliton intensity management problem. Chirped soliton
pulse of Eq. 3 with the constant intensity can be obtained by using Lemma 2
\begin{equation}
P(Z)=-CZ-1;\text{ }D_2(Z)=C/(1+CZ)^2;\text{ }\Gamma (Z)=-C/2/(1+CZ)
\end{equation}
Let us consider some periodical chirped soliton solutions of Eq. 3. Suppose
that the soliton intensity varies periodically as
\begin{equation}
D_2(Z)P^2(Z)=\Theta (Z)=1+\delta \sin ^{2n}Z
\end{equation}
Then soliton solution in the case of n=2 is determined by Eq. 4 with
parameters:
\begin{equation}
D_2(Z)=\Theta (Z)/P^2(Z);\text{ }P(Z)=C-\left[ Z+\delta \left( \frac{3Z}8-%
\frac{\sin 2Z}4+\frac{\sin 4Z}{32}\right) \right]
\end{equation}

\begin{equation}
\Gamma (Z)=\frac 12\frac{\left( 1+\delta \sin ^4Z\right) }{C-\left[ Z+\delta
\left( \frac{3Z}8-\frac{\sin 2Z}4+\frac{\sin 4Z}{32}\right) \right] }+\frac
12\frac{2\text{ }\sin 2Z\text{ }\sin ^2Z}{\left( 1+\delta \sin ^4Z\right) }
\end{equation}
Let us consider some periodical solutions of Eq. 3 in the case of the
linearly dependent parameters of the media. The simplest solution of Eq. 3
in the form of Eq. 6 is:
\begin{eqnarray}
P(Z) &=&\Upsilon (Z)=-\left( 1+\delta \sin ^2Z\right) ;\text{ }N_2(Z)=D_2(Z)=%
\frac{\delta \sin 2Z}{\left( 1+\delta \sin ^2Z\right) ^2};\text{ } \\
\Gamma (Z) &=&\frac \delta 2\frac{\sin 2Z}{\left( 1+\delta \sin ^2Z\right) }
\end{eqnarray}
The next periodical soliton solution  is given by
\begin{equation}
D_2(Z)=N_2(Z)=\cos Z;\text{ }P(Z)=-\frac 1{\left( C-\sin Z\right) };\text{ }%
\Gamma (Z)=\frac{\cos Z}{2\left( C-\sin Z\right) }
\end{equation}

The main soliton features of the solutions given by theorem 1 and theorem 2
were investigated by using direct computer simulations. We have investigated
the interaction dynamics of particle-like solutions obtained, their
soliton-like character was calculated with the accuracy as high as 10$^{-9}$%
. We also have investigated the influence of high-order effects on the
dynamics of dispersion and amplification management. As follows from
numerical investigations elastic character of chirped solitons interacting
does not depend on a number of interacting solitons and their phases. 
Figure \ref{fig02}
shows the computer simulation dynamics of three hyperbolically growing
solitons Eq. 29. NSE solution with periodic dispersion coefficient is shown
in Figure \ref{fig03}. Here the dispersion management function is
\begin{equation}
D_2(Z)=1+\delta \sin ^2(Z)
\end{equation}
and the soliton solution is given by Eqs. 17-18. In Figure \ref{fig03} 
parameters
C=200 and $\delta =-0.9.$ Figure \ref{fig04} represents the two dispersion managed
solitons interaction in the case of equal phases and in Figure \ref{fig05} the
interaction dynamics of two solitons is shown in the case of opposite
phases. Figure \ref{fig06} shows the intensity managed solitons dynamics of the form
presented by Eq. 38. Figures 
\ref{fig07}-\ref{fig09}
show the nonlinear propagation and
interaction of the dispersion and nonlinear managed solitons of Eqs. 42-43.
The main parameters in computer simulations were C=200; $\delta =\pm 0.9$.
Figure \ref{fig10} illustrates 
the dynamics of the fission of the bound states of two
hyperbolically growing solitons Eqs. 29-30 produced by self-induced Raman
scattering effect given by Eqs.2-3. This remarkable fact also emphasize the
full soliton features of solutions discussed. They not only interact
elastically but they can form the bound states and these bound states split
under perturbations. The possibility to find the plethora of soliton
solutions in the case of strong dispersion management is reported in the
recent paper of Zakharov and Manakov \cite{ZakharovVE99}.



%

\newpageAA

\begin{figure}[h] 
\psfig{file=fig01.ps}
\caption{\label{fig01}
Femtosecond soliton spectral trapping effect.
}
\end{figure}

\newpageAA

\begin{figure}[h]
\psfig{file=fig02.ps}
\caption{\label{fig02}
Mutual interaction of three hyperbolically growing chirped 
solitons of Eq. 29 in the case of equal amplitudes and phases.
}
\end{figure}

\newpageAA

\begin{figure}[h]
\psfig{file=fig03.ps}
\caption{\label{fig03}
Evolution of the chirped dispersion managed solitary wave of Eqs. 
17 and 18 as a function of the propagation distance. Dispersion managed 
function is periodic of the form Eq. 45. Input conditions : $C$=200 and 
$\delta$=-0.9.
}
\end{figure}

\newpageAA

\begin{figure}[h] 
\psfig{file=fig04.ps}
\caption{\label{fig04}
Two dispersion managed solitons of Eqs. 17-18 and 45 interaction 
for the case of equal phases. Input conditions: $C$=200 and $\delta$=-0.9.
}
\end{figure}

\newpageAA

\begin{figure}[h]
\psfig{file=fig05.ps}
\caption{\label{fig05}
Two dispersion managed solitons of Eqs. 17-18 and 45 interaction 
for the case of equal phases. Input conditions: $C$=200 and $\delta$=-0.9.
}
\end{figure}

\newpageAA

\begin{figure}[h]
\psfig{file=fig06.ps}
\caption{\label{fig06}
Two intensity managed solitons of Eq. 38 interaction for the case 
of zero initial group velocities.
}
\end{figure}

\newpageAA

\begin{figure}[h]
\psfig{file=fig07.ps}
\caption{\label{fig07}
Evolution of the chirped solitaty wave of Eqs. 39-41 as a function 
of the propagation distance. Input conditions: $\delta$=0.9 and 
group velocity $V_0$=10.
}
\end{figure}

\newpageAA

\begin{figure}[h]
\psfig{file=fig08.ps}
\caption{\label{fig08}
Evolution of the chirped solitary wave of Eqs. 39-41 for the case: 
$\delta$ =-0.8 and group velocity $V_0$=2.0.
}
\end{figure}

\newpageAA

\begin{figure}[h]
\psfig{file=fig09.ps}
\caption{\label{fig09}
Soliton dispersion trapping effect in the presence of the linearly 
dependence between the nonlinearity and GVD parameters.
}
\end{figure}

\newpageAA

\begin{figure}[h]
\psfig{file=fig10.ps}
\caption{\label{fig10}
Decay of high-order hyperbolically growing solitons in the 
presence of third-order dispersion and Raman self-scattering effects.
}
\end{figure}


\begin{thebibliography}{10}

\bibitem{HasegawaA73} 
A. Hasegawa, F. Tappert, ``Transmission of stationary nonlinear optical
pulses in dispersive dielectrical fibers'', Appl. Phys. Lett., v. 23, pp.
142-144, 1973.

\bibitem{MollenauerLF80} 
L.F. Mollenauer, R.G. Stolen, J.P.Gordon,
``Experimental observation of picosecond pulse narrowing and solitons in
optical fibers'', Phys. Rev. Lett., v. 45, pp. 1095-1098, 1980.

\bibitem{ZakharovVE71} 
V.E. Zakharov and A.B. Shabat, ``Exact theory of
two-dimensional self-focusing
and one-dimensional self-modulation of waves in nonlinear media'', Zh. Eksp.
Teor. Fiz., v. 36, pp.61-71, 1971.

\bibitem{TaylorJR92} 
{\it Optical solitons - theory and
experiment}, ed. by J.R. Taylor, Cambridge Univ. Press, 1992

\bibitem{SerkinVN93} 
V.N. Serkin, V.A. Vysloukh and J.R. Taylor, ``Soliton spectral tunneling
effect'', Electron. Lett., v. 29, pp. 12-13, 1993.

\bibitem{DianovEM85} 
E.M. Dianov, A.Ya.
Karasik, P.V. Mamyshev, A.M. Prokhorov, V.N. Serkin,
M.F. Stel'makh and A.A. Fomichev,
``Stimulated Raman conversion of multisoliton pulses in quartz
optical fibers'', JETP Lett., v. 41, pp. 294-297, 1985.

\bibitem{MitschkeFM86} 
F.M. Mitschke and L.F. Mollenauer,
``Discovery of the soliton self-frequency shift'',
Opt. Lett., v. 11, pp. 659-661, 1986.

\bibitem{AfanasjevVV92} 
V.V. Afanasjev, V.N. Serkin and V.A.
Vysloukh, ``Amplification and compression of femtosecond optical solitons in
active fibers'', Sov. Lightwave Commun., v. 2, pp. 35-58, 1992.

\bibitem{KumarS97} 
S. Kumar, A. Hasegawa, ``Quasi-soliton propagation in dispersion managed
optical fibers'', Opt. Lett., v. 22, pp. 372-374, 1997.

\bibitem{HasegawaA98} 
Akira Hasegawa, ``Quasi-soliton for ultra-high speed communications'',
Physica D, v. 123, pp. 267-270, 1998.

\bibitem{KodamaY98} 
Yuji Kodama, ``Nonlinear pulse propagation
in dispersion managed system'', Physica D, V. 123, pp. 255-266, 1998.

\bibitem{MooresJD96} 
John D. Moores, ``Nonlinear compression of chirped solitary waves with and
without phase modulation'', Opt. Lett., v. 21, pp. 555-557, 1996.

\bibitem{KhasilevVY96} 
V.Y. Khasilev, ``Optimal control of all-optical communication soliton
systems'', SPIE Proceedings, v. 2919, pp. 177-188, 1996.

\bibitem{ZakharovVE99} 
V.E.Zakharov and S.V. Manakov, ``On propagation of short pulses in strong
dispersion managed optical lines'', JETP Lett., v.70, pp. 578-582, 1999.

\end{thebibliography}
\end{document}